# Control of spin-orbit torques through crystal symmetry in WTe$_2$/ferromagnet bilayers

D. MacNeill[†1], G. M. Stiehl[†1], M. H. D. Guimaraes[1,2], R. A. Buhrman[3], J. Park[2,4], and D. C. Ralph*[1,2]

1.) Laboratory of Atomic and Solid State Physics, Cornell University, Ithaca, New York 14853, USA.
2.) Kavli Institute at Cornell for Nanoscale Science, Ithaca, New York, 14853, USA.
3.) School of Applied and Engineering Physics, Cornell University, Ithaca, New York 14853, USA.
4.) Department of Chemistry and Chemical Biology, Cornell University, Ithaca, New York 14853, USA.
* Corresponding Author.
† These authors contributed equally to this work.

**Abstract:**

Recent discoveries regarding current-induced spin-orbit torques produced by heavy-metal/ferromagnet and topological-insulator/ferromagnet bilayers provide the potential for dramatically-improved efficiency in the manipulation of magnetic devices. However, in experiments performed to date, spin-orbit torques have an important limitation – the component of torque that can compensate magnetic damping is required by symmetry to lie within the device plane. This means that spin-orbit torques can drive the most current-efficient type of magnetic reversal (antidamping switching) only for magnetic devices with in-plane anisotropy, not the devices with perpendicular magnetic anisotropy that are needed for high-density applications. Here we show experimentally that this state of affairs is not fundamental, but rather one can change the allowed symmetries of spin-orbit torques in spin-source/ferromagnet bilayer devices by using a spin source material with low crystalline symmetry. We use WTe$_2$, a transition-metal dichalcogenide whose surface crystal structure has only one mirror plane and no two-fold rotational invariance. Consistent with these symmetries, we generate an out-of-plane antidamping torque when current is applied along a low-symmetry axis of WTe$_2$/Permalloy bilayers, but not when current is applied along a high-symmetry axis.  Controlling S-O torques by crystal symmetries in multilayer samples provides a new strategy for optimizing future magnetic technologies.

Current-induced torques generated by materials with strong spin-orbit (S-O) interactions are a promising approach for energy-efficient manipulation of nonvolatile magnetic memory and logic technologies[1]. However, S-O torques observed to date are limited by their symmetry so that they cannot efficiently switch the nanoscale magnets with perpendicular magnetic anisotropy (PMA) that are required for high-density applications[2]. S-O torques generated either in conventional heavy metal/ferromagnet thin-film bilayers[3-13] or in topological insulator/ferromagnet bilayers[14,15] are restricted by symmetry to have a particular form[16]: an "antidamping-like" component oriented in the sample plane that is even upon reversal of the magnetization direction, $\hat{m}$, plus an "effective field" component that is odd in $\hat{m}$. The fact that the antidamping torque lies in-plane means that the most efficient mechanism of S-O-torque-driven magnetic reversal for small devices (antidamping switching)[17,18] is available only for magnetic samples with in-plane magnetic anisotropy[8,9], and not PMA samples. S-O torques can also arise from broken crystalline inversion symmetry, even within single layers of ferromagnets[19-22] or antiferromagnets[23], but the antidamping torques that have been measured to date are still limited to lie in the sample plane[21,22,24]. Here we demonstrate that the allowed symmetries of S-O torques in spin source/ferromagnet bilayer samples can be changed by using a spin source material with reduced crystalline symmetry. We generate an out-of-plane antidamping S-O torque when current is applied along a low-symmetry axis of the bilayer. This previously-unobserved form of S-O torque is quenched when current is applied along a high symmetry axis.

As our low-symmetry spin source material, we use the semi-metal WTe$_2$, a layered orthorhombic transition metal dichalcogenide (TMD) with strong S-O coupling[25-29]. TMD materials are attractive for use as sources of S-O torque because they can be prepared as monocrystalline thin films with atomically-flat surfaces down to the single-layer level. They provide a broad palette of crystal symmetries, S-O coupling strengths, and electrical conductivities[30,31]. Other research groups have demonstrated recently the generation of S-O torques in devices made with the TMD MoS$_2$[32], and the Onsager reciprocal process (voltage generation from spin pumping) in $MoS_2$/Al/Co heterostructures[33]. Compared to $MoS_2$, the crystal structure of $WTe_2$ has lower symmetry, with the space group $Pmn2_1$ for bulk $WTe_2$ crystals[34]. In a $WTe_2$/ferromagnet bilayer sample, the screw-axis and glide plane symmetries of this space group are broken at the interface, so that $WTe_2$/ferromagnet bilayers have only one symmetry, a mirror symmetry relative to the bc plane depicted in Fig. 1a. There is no mirror symmetry in the ac plane, and therefore no 180° rotational symmetry about the c-axis (perpendicular to the sample plane).

We fabricate our devices by mechanically exfoliating an artificially-grown crystal of $WTe_2$ onto a high-resistivity oxidized Si wafer, transferring to a high-vacuum sputter system without exposure to air, and sputter depositing 6 nm of Permalloy (Py = $Ni_{81}Fe_{19}$) with a 1 nm protective Al cap (see Methods). The Py magnetic moment is in-plane for all devices studied. The coated $WTe_2$ flakes are patterned into bars by electron-beam lithography and Ar ion milling (Fig. 1b,c), and electrical connection is made by Ti/Pt contact pads. After completion of the device fabrication, we determine the crystallographic axes of each $WTe_2$ flake relative to the bar using polarized Raman spectroscopy (Methods, Supplemental Information). We study only devices in which the active region has minimal surface roughness (<300 pm) and no monolayer steps in the $WTe_2$ as measured by atomic force microscopy. We have measured a total of 15 devices with bars oriented at a variety of alignments to the $WTe_2$ crystal axes and with $WTe_2$ thicknesses ranging from 1.8 nm to 15.0 nm (Supplemental Information Table S1).

To measure the S-O torques produced by our $WTe_2$/Py bilayers, we use the technique of spin-torque ferromagnetic resonance (ST-FMR)[6,21], performed at room temperature. In ST-FMR, an in-plane alternating current is applied through the bilayer at a frequency characteristic of ferromagnetic resonance (here, 5 – 12 GHz). The torques generated by the current excite the magnetic moment away from equilibrium and cause it to precess, creating a time-dependent change in the resistance of the bilayer due to the anisotropic magnetoresistance (AMR) in the ferromagnet. This change in resistance mixes with the alternating current to create a DC voltage across the bar, $V_{mix}$. The circuit used to measure $V_{mix}$ is depicted in Fig. 1c. By sweeping an applied in-plane magnetic field we tune the ferromagnetic resonance through the applied frequency, giving rise to a resonance feature in $V_{mix}$ (Figs. 1d,e). The in-plane ($\tau_\parallel$) and out-of-plane ($\tau_\perp$) torque amplitudes defined in Fig. 1b contribute to the symmetric and antisymmetric parts of the $V_{mix}$ lineshape, respectively. This allows determination of the torque components by fitting $V_{mix}$ as a function of applied magnetic field to a sum of symmetric and antisymmetric Lorentzians (Supplemental Information). The amplitudes of the Lorentzians are related to the two components of torque by[6]:

$$V_S = -\frac{I_{RF}}{2}\left(\frac{dR}{d\phi}\right)\frac{1}{\alpha_G \gamma (2B_0 + \mu_0 M_{eff})}\tau_\parallel, \qquad (1)$$

$$V_A = -\frac{I_{RF}}{2}\left(\frac{dR}{d\phi}\right)\frac{\sqrt{1+\mu_0 M_{eff}/B_0}}{\alpha_G \gamma (2B_0 + \mu_0 M_{eff})}\tau_\perp, \qquad (2)$$

where $R$ is the device resistance, $\phi$ is the angular orientation of the magnetization relative to the direction of applied current in the sample, $dR/d\phi$ is due to the AMR in the Py, $\mu_0 M_{eff}$ is the out-of-plane demagnetization field, $B_0$ is the resonance field, $I_{RF}$ is the microwave current in the bilayer, $\alpha_G$ is the Gilbert damping coefficient and $\gamma$ is the gyromagnetic ratio. In our devices, $\mu_0 M_{eff}$ = 0.7 Tesla and $\alpha_G$ = 0.011 as determined by the ST-FMR resonance frequency and linewidth, respectively, and $R(\phi)$ is measured directly by rotating the magnetic field using a projected-field apparatus.

During a ST-FMR measurement, the applied magnetic field fixes the average angle of the magnetization at a given value, $\phi$. The strengths of the current-induced torques for different angles of the magnetization are related to the symmetries of the device. For example, in a Pt/Py structure, the two-fold rotational symmetry requires that the S-O torque change sign when the magnetization is rotated in-plane by 180°, correspondingly changing the sign of $V_{\text{mix}}$ but maintaining the same magnitude. This is illustrated in Fig. 1d where we plot ST-FMR traces for a Pt(6 nm)/Py(6 nm) bilayer at $\phi$ = 40° and 220°, showing nearly identical lineshapes after multiplying the $\phi$ = 220° trace by -1.

Figure 1e shows the results of the same experiment carried out on a WTe$_2$/Py bilayer with the current applied along the low-symmetry crystal axis of WTe$_2$, parallel to the a-axis (device 1). In this case, we find that $V_{\text{mix}}(40°)$ and $-V_{\text{mix}}(220°)$ differ significantly in both amplitude and shape, indicating that the current-induced torques in the two cases differ in both magnitude and direction. This observation is incompatible with two-fold rotational symmetry, indicating that the current-induced torques are affected by the reduced symmetry of the WTe$_2$ surface.

To analyze this result in more detail, we consider the full angular dependence of the ST-FMR signal as an external magnetic field is used to rotate the direction of the magnetization within the sample plane. In a simple heavy metal/ferromagnet bilayer with no broken lateral symmetries, the current-induced torque amplitudes (due to the spin Hall effect, the Rashba-Edelstein effect, or the Oersted field) have a $\cos(\phi)$ dependence[6,14]. The AMR in Permalloy has an angular dependence that scales as $\cos^2(\phi)$, which enters $V_{\text{mix}}$ as $dR/d\phi \propto \sin(2\phi)$. The product of these two contributions then yields the same angular dependence for the symmetric and antisymmetric ST-FMR components: $V_S = S\cos(\phi)\sin(2\phi)$ and $V_A = A\cos(\phi)\sin(2\phi)$. Our Pt/Py control samples are well described by this behavior (Fig. 2a; the parameter $\phi_0$ accounts for any misalignment between the sample and the electro-magnet, and is typically < 5°).

For our WTe$_2$/Permalloy samples with current along the a-axis, the symmetric component of the ST-FMR signal also has this form (Fig. 2b top panel). The non-zero symmetric component indicates that S-O torques are present in the WTe$_2$/Permalloy bilayer, since the symmetric component corresponds to an in-plane torque and cannot be generated by an Oersted field. However, the more striking result is that the angular dependence of the anti-symmetric component is very different from $\cos(\phi)\sin(2\phi)$ (Fig. 2b bottom panel). The variations in the absolute values of signal amplitudes reflect the broken symmetries of the WTe$_2$ surface: the absence of mirror symmetry in the ac plane (corresponding to $\phi \rightarrow 180° - \phi$, since $\hat{m}$ is a pseudovector) and the absence of twofold rotational symmetry about the c-axis ($\phi \rightarrow 180° + \phi$). This result indicates the existence of a source of out-of-plane torque not previously observed in any S-O torque experiment.

The unusual angular dependence we measure for the antisymmetric ST-FMR signal with current applied along the a-axis can be well fit by the simple addition of a term proportional to $\sin(2\phi)$:

$$V_A(\phi) = A\cos(\phi)\sin(2\phi) + B\sin(2\phi) \tag{3}$$

where $A$ and $B$ are constants independent of the field angle (see the solid line in Fig. 2b bottom panel). To quantitatively translate the measured angular dependence of $V_S$ and $V_A$ to torques, we can use Eqs. (1) and (2) to remove the contribution from the angular dependence of the AMR. The fits in Figs. 2b then correspond to angular dependences for the in-plane and perpendicular torque amplitudes of the form

$$\tau_\parallel(\phi) = \tau_S \cos(\phi), \tag{4}$$

$$\tau_\perp(\phi) = \tau_A \cos(\phi) + \tau_B, \tag{5}$$

where $\tau_S$, $\tau_A$, and $\tau_B$ are independent of $\phi$. The terms proportional to $\cos(\phi)$ are the usual terms observed previously, and in the Pt/Py control samples. The new term ($\tau_B$) corresponds to an out-of-plane torque that is independent of the in-plane magnetization orientation; *i.e.*, it is even in $\hat{m}$ and therefore an antidamping-like torque. It is consistent with predictions[35] that broken lateral mirror symmetry can allow an out-of-plane torque of the form $\vec{\tau}_{AD} \propto \hat{m} \times (\hat{m} \times \hat{c})$. That an out-of-plane antidamping-like torque with the form of $\tau_B$ could exist has also been discussed in an analysis of the allowed symmetries for S-O torques in GaMnAs /Fe samples[24], but this torque has not previously been identified in experiment.

In commonly studied bilayer systems without any broken in-plane symmetries, a linear-in-current out-of-plane torque that is independent of the in-plane magnetization angle cannot exist by symmetry. For example, the presence of a twofold rotation disallows $\tau_B$. In samples with twofold rotational symmetry, rotating the sample by 180° is equivalent to changing the sign of an in-plane current without changing the sign of $\tau_B$, which violates the linear-in-current requirement. However, WTe$_2$/Py bilayers do not have two-fold rotational symmetry. The only symmetry in our WTe$_2$/Py bilayers is the bc plane mirror, $\sigma_v(\mathrm{bc})$. The effect of $\sigma_v(\mathrm{bc})$ on a WTe$_2$/Py bilayer with current flowing along the a-axis is illustrated in Fig. 2c. Both the out-of-plane torque (a pseudovector) and the current change sign under $\sigma_v(\mathrm{bc})$, $\tau_B \to -\tau_B$ and $I \to -I$, which is the expected behavior for a current-generated S-O torque: the sign of torque must change with the sign of the current. A torque with the symmetry of $\tau_B$ is therefore allowed for WTe$_2$/Py bilayers with current along the a-axis.

We observe that $\tau_B$ goes to zero when the current is applied parallel to the b-axis of WTe$_2$. Figure 3a shows the antisymmetric ST-FMR component $V_A$ (red circles) as a function of $\phi$ for device 2, in which the current is applied along the b-axis. The angular fit to Eq. (3) yields a value of *B* equal to 0 within experimental uncertainty. This result is again consistent with the symmetries of the WTe$_2$ surface layer. When the mirror symmetry operation $\sigma_v(\mathrm{bc})$ is applied in this case (Fig. 3b), the out-of-plane torque is inverted but the current is not, and therefore an out-of-plane component of torque that is independent of the in-plane magnetization angle is forbidden by symmetry. Higher-order angular terms [namely $\tau_\perp \propto \cos(n\phi)$ with *n* odd or $\tau_\perp \propto \sin(m\phi)$ with *m* even] are symmetry-allowed for current along the b-axis, and can be included in fits of $V_A$ versus $\phi$ to improve the quantitative agreement (Supplemental Information). We also continue to observe a nonzero symmetric ST-FMR signal when the current is aligned with the b-axis, which has the same functional form as the symmetric ST-FMR signal in the devices with current along the a-axis (Supplemental Information).

We further investigated the symmetry dependence of $\tau_B$ by studying devices with different angles, $\phi_{a-I}$, between the a-axis of the WTe$_2$ and the applied current direction. We fabricated 15 devices with different $\phi_{a-I}$ and performed full angle-dependent ST-FMR measurements on each in order to extract $A$, $B$ and $S$ (Supplemental Information Table S1). Figure 4a shows the ratio of $\tau_B / \tau_A$ at a given frequency ($f$ = 9 GHz) as a function of $\phi_{a-I}$ for these different devices. We consistently see that the ratio of $\tau_B / \tau_A$ is large when current is aligned with the a-axis, and is gradually quenched as the projection of the current along the b-axis grows. This provides strong additional evidence that the observed magnetization-independent out-of-plane torque is correlated with the symmetries present in the WTe$_2$ crystal.

The dependence of the measured torques on WTe$_2$ thickness provides insight into the mechanism of torque generation. If the torques arise through a bulk mechanism, a clear thickness dependence should be expected; however, if the torques are generated by an interface effect they should not depend on WTe$_2$ thickness. Figure 4b shows the dependence of the torque ratios on WTe$_2$ thickness for devices that have current along the a-axis. Neither $\tau_B / \tau_A$ nor $\tau_S / \tau_A$ show any significant thickness dependence. This requires $\tau_S$, $\tau_A$, and $\tau_B$ to either all have the same thickness dependence or have no definite thickness dependence. However, as $\tau_B$ can originate only at the interface due to symmetry, this suggests that generation of all three torques arise from an interfacial effect in the WTe$_2$/Py bilayer.

The strength of the individual components of torque can be determined quantitatively from Eqs. (1) and (2), using independently-measured values of the resistance as a function of magnetization angle ($dR/d\phi$) and the transmitted and reflected microwave power ($S_{21}$ and $S_{11}$) in order to determine $I_{RF}$ (Supplemental Information). We will express these strengths as torque conductivities ($\sigma_S$, $\sigma_A$, $\sigma_B$; torques per unit area per unit electric field) because the electric field applied across the device can be determined accurately, while the division of current density flowing in the different layers has larger uncertainties. We find for current along the a-axis that $\sigma_S = (8 \pm 2) \times 10^3$ ($\hbar/2e$) ($\Omega$m)$^{-1}$, $\sigma_A = (9\pm3) \times 10^3$ ($\hbar/2e$) ($\Omega$m)$^{-1}$, and $\sigma_B = (3.6 \pm 0.8) \times 10^3$ ($\hbar/2e$) ($\Omega$m)$^{-1}$, where the uncertainties give the standard deviation across our devices (Supplemental Information).

We find it interesting that although a broken lateral mirror symmetry should also allow additional terms for the in-plane S-O torque when current is applied along the a-axis, for example an effective-field torque of the form $\propto \hat{m} \times \hat{c}$ [35], we detect no such contributions. A term $\propto \hat{m} \times \hat{c}$ would add a $\phi$-independent contribution to Eq. (4), $\tau_\parallel(\phi) \rightarrow \tau_S \cos(\phi) + \tau_T$ that would cause the absolute values of the amplitudes for the symmetric part of the ST-FMR resonance (Fig. 2c) to be asymmetric under the operations $\phi \rightarrow 180° - \phi$ and $\phi \rightarrow 180° + \phi$. We can set a limit for our devices that $|\tau_T| \leq 0.05 \tau_S$. Our results are therefore opposite a report about S-O torques in "wedge" samples[35], which claimed that the breaking of lateral mirror symmetry by the wedge structure generated an effective field torque $\propto \hat{m} \times \hat{c}$, but no out-of-plane antidamping torque. We question whether the extremely small thickness gradient in ref. 35 (a difference in average thickness of < 0.5 picometers, or 0.002 of an atom, between the two sides of a 20-μm-wide sample) actually provides a meaningful breaking of structural mirror symmetry.

We note one additional consequence of strong S-O coupling at the WTe$_2$/Py interface – the magnetic anisotropy easy axis of the Py is determined by the crystal lattice of the WTe$_2$. The magnetic anisotropy can be determined from our ST-FMR data via the $\phi$ dependence of the magnetic resonance frequency and by direct AMR measurements (Supplemental Information). Regardless of the orientation of the sample channel with respect to the WTe$_2$ crystal lattice, we find that the magnetic easy axis is always parallel to the b-axis of WTe$_2$. The effective anisotropy field in different devices ranges from 4.9 to 17.3 mT for 6 nm of Py (Supplemental Information Table S1).

In summary, we have demonstrated that it is possible to generate an out-of-plane antidamping-like S-O torque in spin-source/ferromagnet bilayers by using a spin-source material whose surface crystal structure has a broken lateral mirror symmetry. This is important as it provides a strategy for achieving efficient manipulation of magnetic devices with perpendicular magnetic anisotropy. Compared to in-plane-magnetized devices, PMA devices are of interest because they can be scaled to smaller sizes and higher density while maintaining thermal stability. PMA devices can be switched much more efficiently using an out-of-plane antidamping torque, $\tau_{AD}$, compared to an effective field torque, $\tau_{FL}$, since the effective field torque required for switching in the macrospin limit is $|\tau_{FL}| \approx \gamma H_{anis}$, where $H_{anis}$ is the anisotropy field, while the antidamping torque required is much smaller: $|\tau_{AD}| \approx \alpha_G \gamma H_{anis}$, with $\alpha_G \approx 0.01$.[18] Previously, because S-O torques could generate an antidamping-like component only in the sample plane, they have been incapable of switching PMA devices by this efficient antidamping process[36] – an in-plane anti-damping torque switches PMA devices through a mechanism involving domain nucleation and domain-wall propagation[37-41] that becomes inefficient at small size scales[2]. Our results therefore suggest a strategy, based on control of broken crystal symmetry in materials with strong S-O coupling, that has the potential to enable efficient antidamping switching of PMA memory and logic devices at the 10's of nm size scale.

**Methods:**

**Device Fabrication**

Our device fabrication starts with high quality artificially-grown crystals of $WTe_2$ (HQ Graphene) which we exfoliate in a nitrogen-rich environment onto a high-resistivity silicon wafer with 1 μm of thermal oxide. This results in the deposition of single-crystal flakes up to 40 μm in lateral extent. The exfoliated samples are transferred into a high-vacuum sputtering system without exposure to air. To minimize damage to the $WTe_2$ flakes, we use grazing-angle magnetron sputtering to deposit 6 nm of Permalloy (Py=$Ni_{81}Fe_{19}$) onto the $WTe_2$. The Py deposition rates are kept below 0.2 Å/s and are performed in an ambient Ar background pressure of 4 mtorr. We then deposit a protective aluminum oxide cap *in situ* onto the $WTe_2$ / Py bilayer by sputter deposition of 1 nm aluminum which is subsequently oxidized in a dry $N_2/O_2$ mixture.

After deposition of the ferromagnet and aluminum oxide cap, we use optical contrast and atomic force microscopy (AFM) to select $WTe_2$ flakes for further study. Flakes are chosen to ensure an active region with homogenous thickness (i.e. no monolayer steps or tape residue) and minimal roughness (typically < 300 pm RMS). An AFM image of a typical $WTe_2$ / Py bilayer prior to patterning is shown in the Supplemental Information.

The $WTe_2$ / Py bilayers are patterned into bars of width 3-4 μm. The bars are defined via Ar ion milling, using either a hard mask (silicon or aluminum oxide) or an e-beam exposed PMMA/HSQ bilayer. After etching, another step of e-beam lithography is used to make electrical contact to the bars with Ti(5 nm)/Pt (50 nm) contact pads, which have a ground-signal-ground geometry compatible with microwave probes (Fig. 1c). The active region between the contacts is 3-6 μm long. For hard-mask devices, an additional reactive-ion etching (RIE) or wet etch step is used prior to the Ti/Pt deposition to remove the mask in the contact region.

The Py resistivity in our devices is (100 ± 20) μΩ cm. The WTe$_2$ bulk resistivity value is (380 ± 10) μΩ cm with the current flowing along the a-axis and is likely higher in thinner flakes[42]. The resistivity of WTe$_2$ is anisotropic, and we find the resistivity with current flowing along the a-axis to be 1.4 – 2 times larger than the resistivity for current flowing along the b-axis.

**Measurements**

For ST-FMR measurements a microwave current at a fixed frequency (ranging from 5-12 GHz) is applied to the WTe$_2$/Py bilayer through a bias T (Fig. 1c). The amplitude of the microwave current is modulated at kHz frequencies, which allows a lock-in measurement of the DC mixing voltage ($V_{\text{mix}}$) at the inductive side of the bias T. In-plane magnetic fields are applied using a projected field electromagnet which can be rotated 360° about the vertical axis. Magnetic fields are swept from 0.24 T to 0 T to drive the Py through its resonance condition. The transmission and reflection coefficients for the RF cabling and devices, respectively, are determined by calibrated vector network analyzer measurements in the relevant frequency range (5-12 GHz). Measurements of the AMR in the Py are made in a constant 0.08 T field using a Wheatstone bridge and a lock-in amplifier.

**Determination of the WTe$_2$ crystal axes**

The crystal axes of WTe$_2$ are determined by polarized Raman measurements using a Renishaw inVia confocal Raman microscope with a linearly polarized 488 nm wavelength excitation and a co-linear polarizer placed in front of the spectrometer entrance slit. The WTe$_2$ sample is positioned such that the light is incident normal to the sample surface and the excitation electric field is in the sample plane. Previous calculations and measurements have shown that the 165.7 cm$^{-1}$ (P6) and 211.3 cm$^{-1}$ (P7) Raman peaks of WTe$_2$ are sensitive to the alignment of the electric field and the crystal axes[43]. We measure the intensity of the P6 and P7 Raman peaks as a function of angle by rotating the sample from 0° to 180° in steps of 10° or 20° (keeping the electric field in the sample plane). The angle for which the ratio of peak intensities, P6/P7, is maximized identifies the a-axis, allowing determination of the angle between the a-axis and current direction, $\phi_{a-I}$ (Supplementary Information).

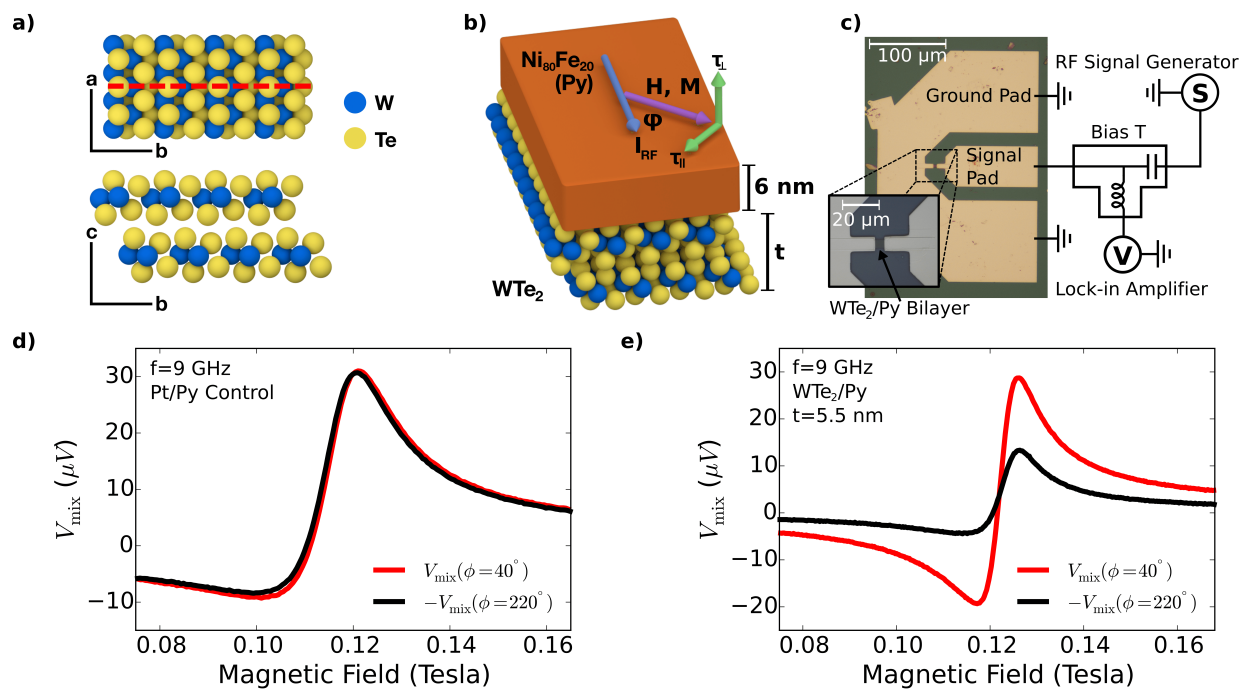

**Fig. 1. Sample geometry and sample ST-FMR results**. (a) Crystal structure near the surface of WTe$_2$. The surface possesses mirror symmetry with respect to the bc plane (dashed line), but not with respect to the ac plane, and therefore it is also not symmetric relative to a 180° rotation about the c-axis. (b) Schematic of the bilayer WTe$_2$/Permalloy sample geometry. (c) Optical images of the sample geometry including contact pads, with the circuit used for spin-torque FMR measurements. (d) ST-FMR resonances for Pt(6 nm)/Py(6 nm) control samples, with the magnetization oriented at 40° and 220° relative to the current direction. The applied microwave power is 0 dBm. (e) ST-FMR resonances for a WTe$_2$(5.5 nm)/Py(6 nm) sample with current applied along the a-axis, with the magnetization oriented at 40° and 220° relative to the current direction. The applied microwave power is 5 dBm.

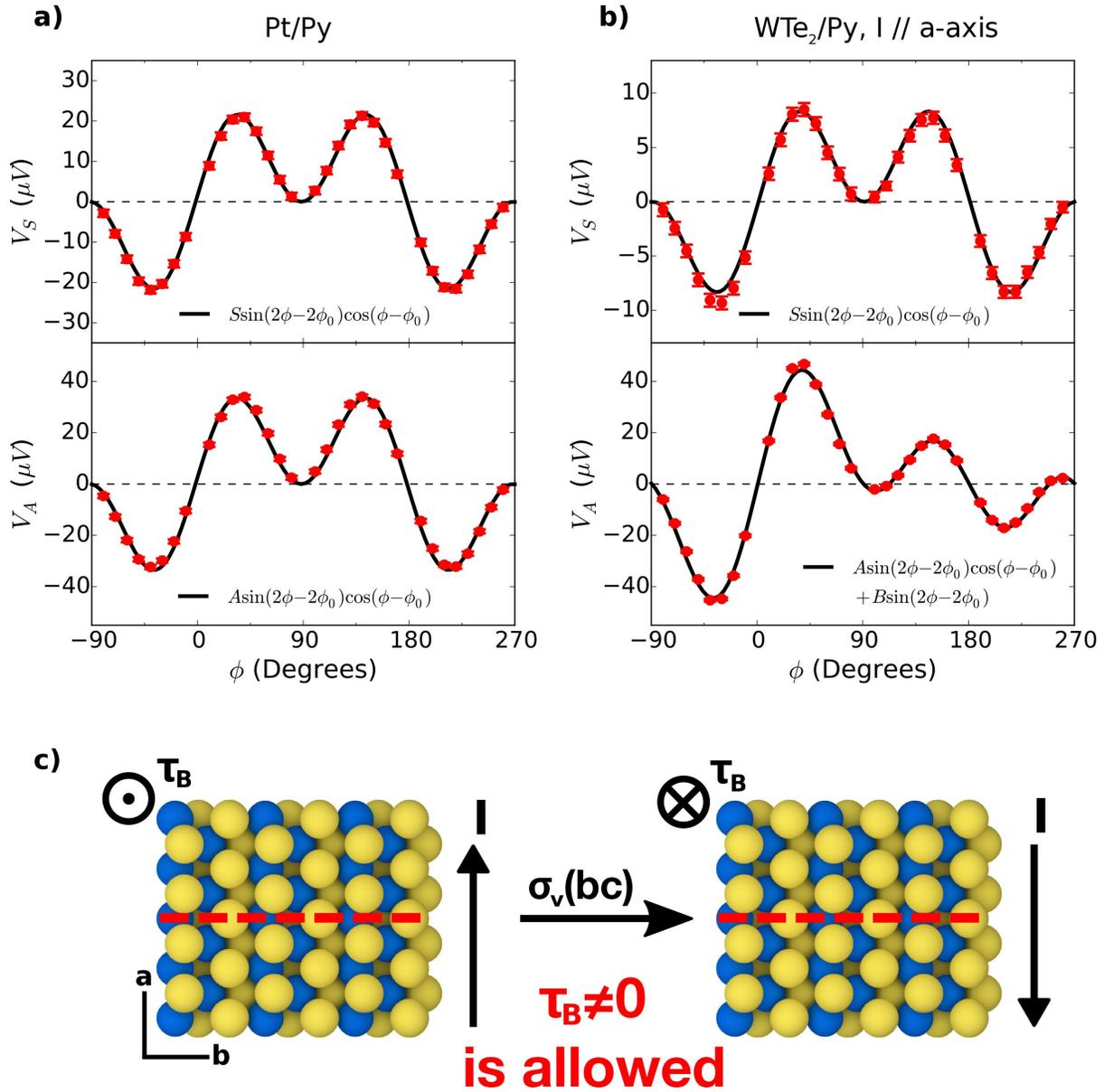

**Fig. 2. Angular dependence of ST-FMR signals.** (a) Symmetric and antisymmetric ST-FMR resonance components for a Pt(6 nm)/Py (6 nm) control sample as a function of in-plane magnetic-field angle. The microwave frequency is 9 GHz and the applied microwave power is 0 dBm. The parameter $\phi_0$ accounts for any misalignment between the sample and the magnet. (b) Symmetric and antisymmetric ST-FMR resonance components for a $WTe_2$(5.5 nm)/Py (6 nm) device (device 1) as a function of in-plane magnetic-field angle, with current applied parallel to the a-axis. The microwave frequency is 9 GHz and the applied microwave power is 5 dBm. (c) Illustration that a magnetization-independent, linear-in-current out-of-plane S-O torque is allowed by symmetry for current applied along the a-axis of a $WTe_2$/Py bilayer.

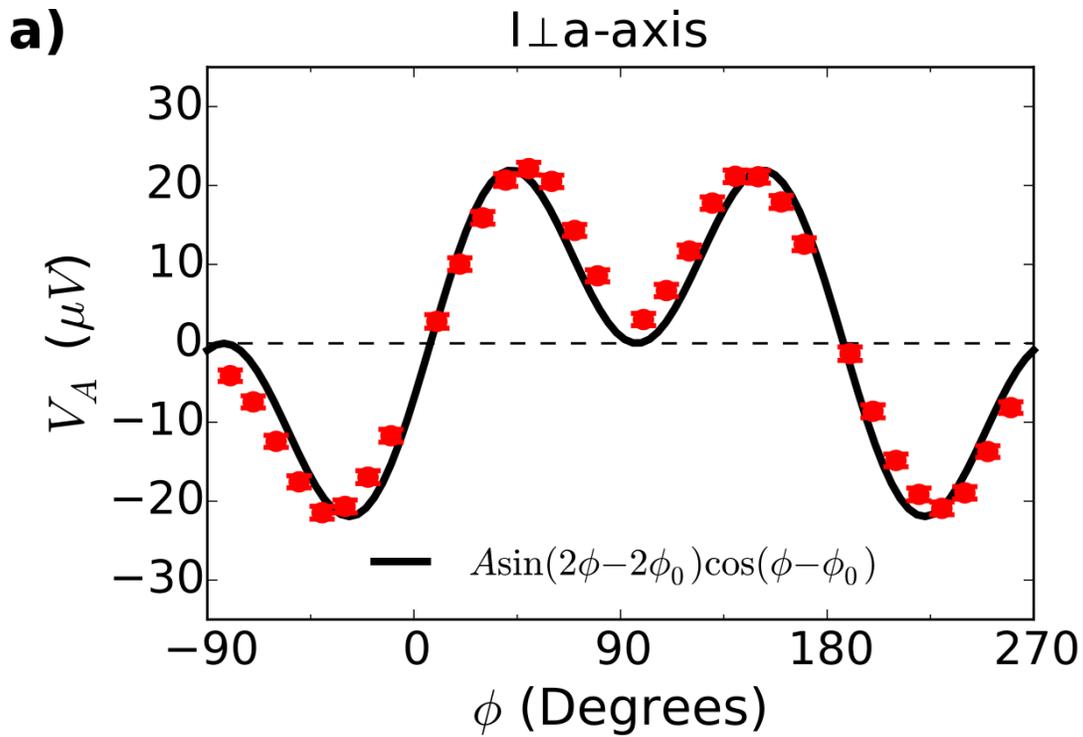

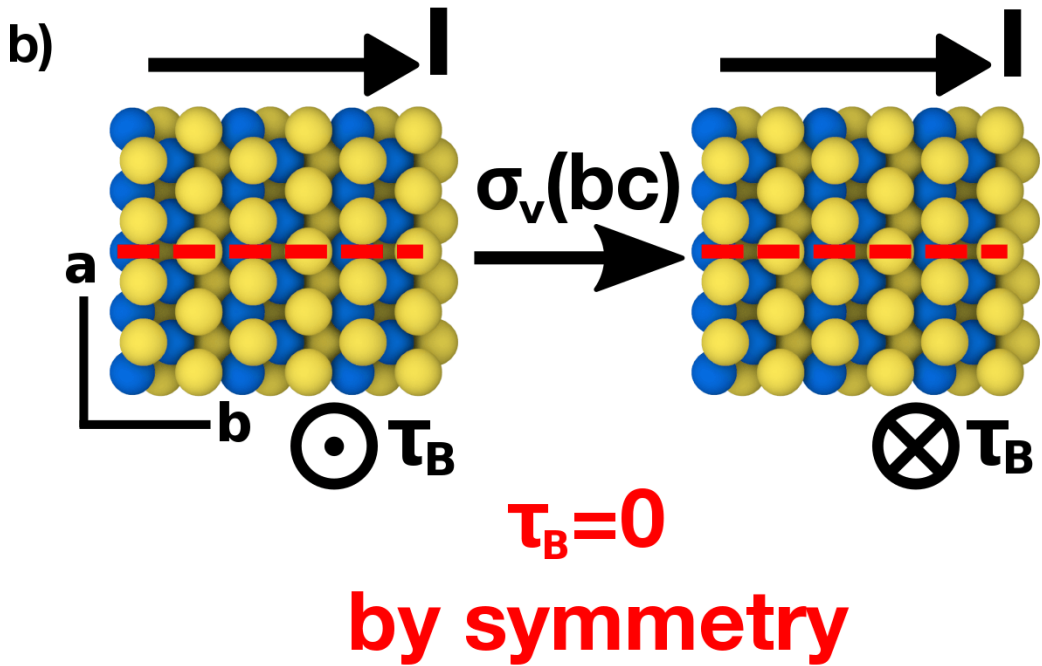

**Fig. 3. Spin-orbit torque for current along the high-symmetry WTe₂ b-axis.** (a) Antisymmetric ST-FMR resonance component for a WTe$_2$(15 nm)/Py (6 nm) device (device 2) as a function of in-plane magnetic-field angle, with current applied parallel to the b-axis. The microwave frequency is 9 GHz and the applied microwave power is 5 dBm. (b) Illustration that a magnetization-independent, linear-in-current out-of-plane S-O torque is forbidden by symmetry for current applied along the b-axis.

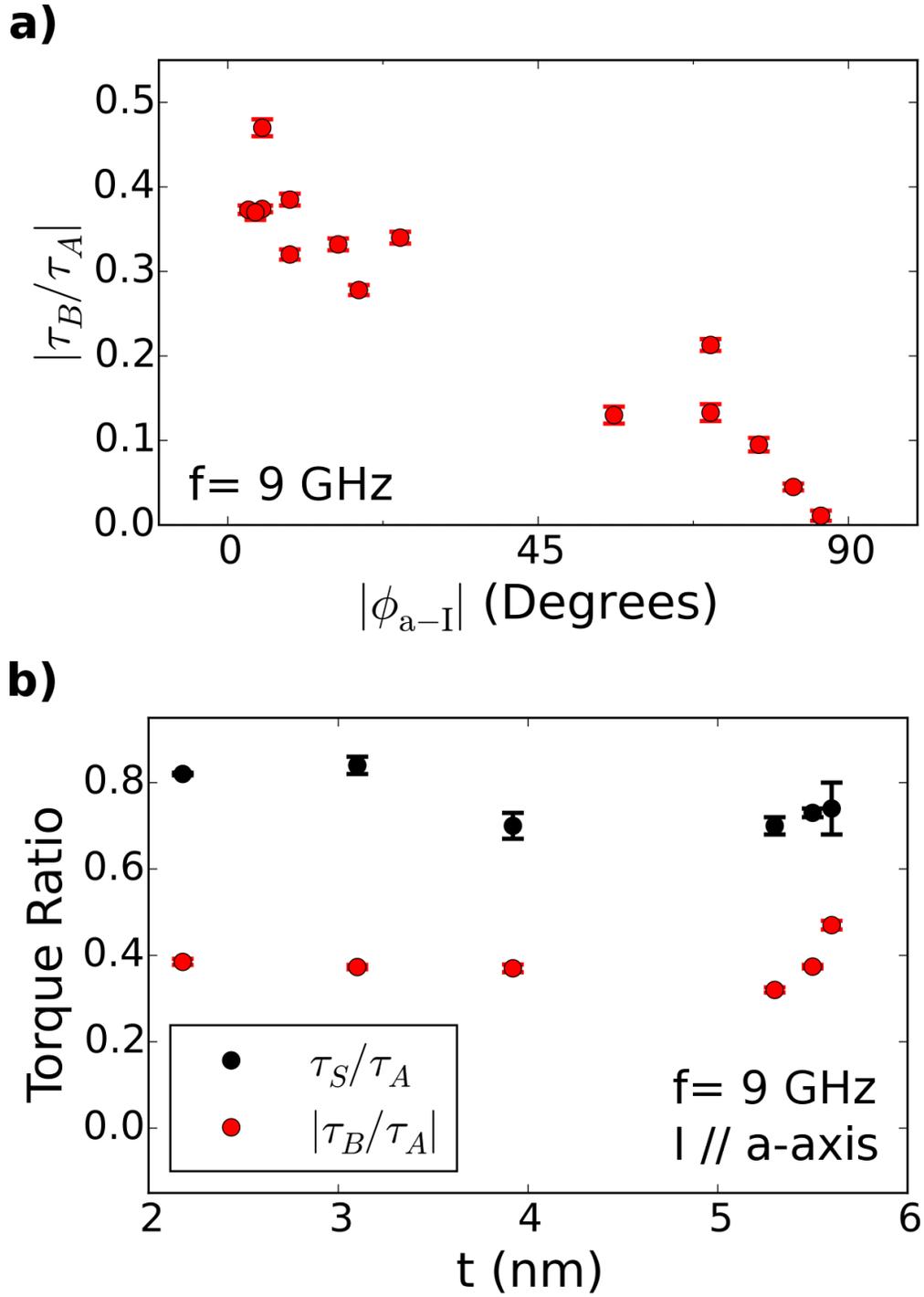

**Fig 4. Dependence of the spin-orbit torques on the angle of applied current and the thickness of WTe₂.** (a) Ratio of the out-of-plane antidamping torque $\tau_B$ to the out-of-plane effective-field torque $\tau_A$ as a function of the angle between the a-axis and the applied current. (b) Torque ratios as a function of the thickness of the WTe₂ layer for current applied along the a-axis. Here $\tau_S$ is the in-plane current-induced torque.

Acknowledgements:

We thank N. D. Reynolds for experimental assistance, R. De Alba for help with the graphics, and G. D. Fuchs and P. G. Gowtham for comments on the manuscript. This work was supported by the National Science Foundation (DMR-1406333) and the Army Research Office (W911NF-15-1-0447). G.M.S. acknowledges support by a National Science Foundation Graduate Research Fellowship under Grant No. DGE-1144153. M.H.D.G. acknowledges support by the Netherlands Organization for Scientific Research (NWO 680-50-1311) and the Kavli Institute at Cornell for Nanoscale Science. This work made use of the Cornell Nanoscale Facility, which is supported by the NSF (ECCS-1542081) and also the Cornell Center for Materials Research Shared Facilities, which are supported through the NSF MRSEC program (DMR-1120296).

Author Contributions:

DJM, GMS, MHDG and DCR conceived of the idea for the experiment. DJM performed the sample fabrication. GMS made the measurements. DJM and GMS performed the analysis with help from MHDG, RAB, JP and DCR. DJM, GMS and DCR wrote the manuscript and all authors contributed to the final version.

Competing Financial Interests:

The authors declare no competing financial interests.

# Control of spin-orbit torques through crystal symmetry in WTe₂/ferromagnet bilayers

D. MacNeill[†1], G. M. Stiehl[†1], M. H. D. Guimaraes[1,2], R. A. Buhrman[3], J. Park[2,4], and D. C. Ralph[*1,2]

1.) Laboratory of Atomic and Solid State Physics, Cornell University, Ithaca, New York 14853, USA.
2.) Kavli Institute at Cornell for Nanoscale Science, Ithaca, New York, 14853, USA.
3.) School of Applied and Engineering Physics, Cornell University, Ithaca, New York 14853, USA.
4.) Department of Chemistry and Chemical Biology, Cornell University, Ithaca, New York 14853, USA.
* Corresponding Author.
† These authors contributed equally to this work.

## Supplemental Information

**Supplementary Note 1: Analysis of ST-FMR measurements**

We model the ST-FMR measurements by using the Landau-Lifshitz-Gilbert-Slonczewski (LLGS) equation to calculate the precessional dynamics of the magnetization direction, $\hat{m}(t)$, in the macrospin approximation in response to the in-plane and out-of-plane torque amplitudes, $\tau_\parallel(\phi)$ and $\tau_\perp(\phi)$ as defined in the main text[S1,S2]. This determines the ST-FMR mixing voltage as

$$V_{\text{mix}} = \left\langle I(t) R[\hat{m}(t)] \right\rangle_t = V_S \frac{\Delta^2}{\left(B_{\text{app}} - B_0\right)^2 + \Delta^2} + V_A \frac{\Delta\left(B_{\text{app}} - B_0\right)}{\left(B_{\text{app}} - B_0\right)^2 + \Delta^2} \tag{S1}$$

where $B_{\text{app}}$ is the applied magnetic field, $B_0$ is the applied magnetic field at ferromagnetic resonance, and $\Delta$ is the linewidth. The $\hat{m}(t)$ dependence of the device resistance, $R$, arises from the anisotropic magnetoresistance (AMR) of the ferromagnet permalloy. We determine the symmetric and antisymmetric amplitudes, $V_S$ and $V_A$, by fitting Eq. S1 to measurements of the mixing voltage as a function of applied magnetic field. These amplitudes are related to the torque amplitudes $\tau_\parallel$ and $\tau_\perp$ by Eqs. 1 and 2 in the main text. We note that $\tau_\parallel$ and $\tau_\perp$ are normalized by the total angular momentum of the magnet, and so have dimensions of frequency. We determine torque ratios from the ratio of Eqs. 1 and 2, together with measured values for $B_0$ and $M_{\text{eff}}$. We obtain the value of $B_0$ via fits of the resonance lineshape to Eq. S1, and we estimate $M_{\text{eff}}$ from the frequency dependence of $B_0$ using the Kittel formula $2\pi f = \gamma \sqrt{B_0(B_0 + M_{\text{eff}})}$. As we discuss in Supplementary Note 2, $B_0$ and $M_{\text{eff}}$ depend on $\phi$ due to the in-plane magnetic anisotropy of our samples. For our analysis we use angle-averaged values for these quantities; the error in doing so is less than 5% due to the small degree of angular variation.

To obtain quantitative measurements of the individual torque components using Eq. 1 or Eq. 2 (*i.e.* not just their ratios), it is also necessary to determine $\alpha_G$, $R(\phi)$, and $I_{\text{RF}}$. The Gilbert damping $\alpha_G$ is estimated from the frequency dependence of the linewidth via $\Delta = 2\pi f \alpha_G / \gamma + \Delta_0$, where $\Delta_0$ is the inhomogeneous broadening. To obtain the AMR we measure the device resistance as a function of a rotating in-plane magnetic field (with magnitude 0.08 T) applied via a projected-field magnet. Fitting these data to $R_0 + \Delta R \cos^2(\phi - \phi_0)$ allows calculation of $dR/d\phi$ (Fig. S1). To measure the RF current $I_{RF}$, we use a vector network analyzer to estimate the reflection coefficients of our devices ($S_{11}$) and the transmission coefficient of our RF circuit ($S_{21}$). These calibrations allow calculation of the RF current flowing in the device as a function of applied microwave power and frequency:

$$I_{RF} = 2\sqrt{1mW \cdot 10^{\frac{P_{source}(dBm) + S_{21}(dBm)}{10}} (1-|\Gamma|)^2 \Big/ 50} , \tag{S2}$$

where $P_{source}$ is the power sourced by the microwave generator and $\Gamma$ is given by

$$\Gamma = 10^{\frac{S_{11}(dBm)}{20}}. \tag{S3}$$

The torque conductivity, defined as the angular momentum absorbed by the magnet per second per unit interface area per unit electric field, provides an absolute measure of the torques produced in a spin source/ferromagnet bilayer independent of geometric factors. For a torque $\tau_K$ (where K = one of the A, B, S, or T indices for the torque components defined in the main text) we calculate the corresponding torque conductivity via

$$\sigma_K = \frac{M_S l w t_{magnet}}{\gamma} \frac{\tau_K}{(lw)E} = \frac{M_S l t_{magnet}}{\gamma} \frac{\tau_K(1-\Gamma)}{I_{RF} \cdot 50(1+\Gamma)} , \tag{S4}$$

where $M_S$ is the saturation magnetization, $E$ is the electric field, $l$ and $w$ are the length and width of the WTe$_2$/permalloy bilayer, and $t_{magnet}$ is the thickness of the permalloy. The factor $M_S l w t_{magnet}/\gamma$ is the total angular momentum of the magnet, which converts the normalized torque into units of angular momentum per second. Due to the unavailability of mm-scale WTe$_2$/permalloy bilayers, we are unable to measure $M_S$ directly via magnetometry, and instead approximate $M_S \approx M_{eff}$, which we have found to be accurate in other permalloy bilayer systems[2].

**Supplementary Note 2: Determination of in-plane magnetic anisotropy**

Figure S2 shows the magnetic field at ferromagnetic resonance as a function of the in-plane magnetization angle for Devices 1 and 2. For Device 1 the current flows nearly parallel to the a-axis ($\phi_{a-I} = -3°$), and for Device 2 it is nearly parallel to the b-axis ($\phi_{a-I} = 86°$). The data from both samples indicate the presence of a uniaxial magnetic anisotropy within the sample plane, with an easy axis along the b-axis of the WTe$_2$. The angular dependence of the resonance field is described well by the form

$$B_0 = B_{Kittel} - B_A \cos(2\phi - 2\phi_{Easy-I} - 2\phi_0) \tag{S5}$$

where $B_A$ is the in-plane anisotropy field, related to the anisotropy energy $K_A$ via $B_A = 2\mu_0 K_A/M_s$, $B_{Kittel}$ is the resonance field without any in-plane anisotropy, $\phi_{Easy-I}$ is the angle from the current direction to the magnetic easy-axis and $\phi_0$ is the angular misalignment extracted from the angular dependence of the mixing voltage as discussed in the main text. This equation also assumes $B_A$, $B_{Kittel} \ll M_{eff}$ which are valid approximations for our experiment. We find values for $B_A$ of 7 mT and 15 mT for Device 1 and Device 2, respectively. We observe no unidirectional component to the magnetic anisotropy.

We performed similar fits for all of the devices listed in Table S1 (Supplementary Note 3). In all cases the magnetic easy axis was along the b-axis within experimental uncertainty; i.e. $\phi_{a\text{-}I} = \phi_{Easy\text{-}I} + 90°$. Over all of our devices we find $B_A$ to be in the range 4.9-17.3 mT. Some, but likely not all, of the device-to-device variation may be explained by differences in the sample shape.

**Supplementary Note 3: Data from additional devices**

In Table S1, we provide device parameters, torque ratios, and magnetic anisotropy parameters for 15 WTe$_2$/permalloy bilayers, and a Pt/permalloy control device. In Fig. S3, we plot $V_S$ and $V_A$ as a function of $\phi$ for four devices, along with fits to $S\sin(2\phi - 2\phi_0)\cos(\phi - \phi_0)$ and $\sin(2\phi - 2\phi_0)[B + A\cos(\phi - \phi_0)]$ for the symmetric and antisymmetric data respectively. The sign of the parameter $B$ varies apparently randomly between devices. This is to be expected because Raman scattering does not allow us to distinguish between the $\hat{b}$ and $-\hat{b}$ directions, which are physically distinct for the WTe$_2$ surface crystal structure (a consequence of broken two-fold rotational symmetry). Essentially, the sign of $B$ depends on whether the positive $\hat{b}$ direction lies along $0 < \phi < 180°$ or $180° < \phi < 360°$. Since interchanging the ground and signal leads rotates the definition of $\phi$ by 180°, the sign of $B$ is determined by the decision of which end of the bilayer is connected to the signal lead.

We carried out calibrated torque conductivity measurements (*i.e.*, using a vector network analyzer to determine $I_{RF}$ as discussed in Supplementary Note 1) for 11 of our devices. The device-averaged torque conductivities for devices with current applied along the a-axis are reported in the main text. The torque conductivity data from all 11 devices is summarized in Fig. S4. In Fig. S4a and Fig. S4b we plot $\sigma_S$ and $\sigma_A$ respectively as a function of thickness. In Fig. S4c we plot $|\sigma_B|$ as a function of thickness for the subset of the 11 devices where current is applied along the a-axis, and in Fig. S4d we plot $|\sigma_B|$ as a function of $|\phi_{a\text{-}I}|$ for all 11 devices.

**Supplementary Note 4: Symmetry analysis for current generated torques.**

The torques acting on an in-plane magnetization can be written as $\vec{\tau}_\parallel(\hat{m}, E) = \tau_\parallel(\phi, E)\hat{m} \times \hat{c}$ and $\vec{\tau}_\perp(\hat{m}, E) = \tau_\perp(\phi, E)\hat{c}$, where we have explicitly included the dependence of the torques on the electric field, $E$, in the bilayer. These expressions are generic, since $\hat{m} \times \hat{c}$ and $\hat{c}$ are unit vectors forming a basis for the vectors perpendicular to $\hat{m}$. The scalar pre-factors, $\tau_\parallel(\phi, E)$ and $\tau_\perp(\phi, E)$, can be Fourier expanded:

$$\tau_\parallel(\phi, E) = E(S_0 + S_1\cos\phi + S_2\sin\phi + S_3\cos 2\phi + S_4\sin 2\phi + S_5\cos 3\phi + \ldots)$$
$$\tau_\perp(\phi, E) = E(A_0 + A_1\cos\phi + A_2\sin\phi + A_3\cos 2\phi + A_4\sin 2\phi + A_5\cos 3\phi + \ldots)$$ (S6)

First, we consider the case of an electric field applied along the WTe$_2$ crystal a-axis. In this case, applying the $\sigma_v(\text{bc})$ symmetry operation to the device flips the direction of the electric field (since $\vec{E}$ is a vector perpendicular to the bc plane) and reverses the component of the magnetization perpendicular to the a-axis (since $\hat{m}$ is a pseudovector). This is equivalent to the transformations $\phi \rightarrow -\phi$ and $E \rightarrow -E$.

The torques must also transform as pseudovectors under $\sigma_v(\text{bc})$, which constrains the dependence of $\tau_\parallel(\phi, E)$ and $\tau_\perp(\phi, E)$ on $\phi$ and $E$. The nature of these constraints can be understood by re-writing $\tau_\perp(\phi, E) = \hat{c} \cdot \vec{\tau}_\perp$ and $\tau_\parallel(\phi, E) = (\hat{m} \times \hat{c}) \cdot \vec{\tau}_\parallel$. Since $\hat{c}$ is a vector and $\vec{\tau}_\perp$ is a pseudovector, $\hat{c} \cdot \vec{\tau}_\perp$ transforms as a pseudoscalar (*i.e.* changes sign under inversion and mirror operations but is invariant under rotations) as the dot product of a vector and a pseudovector is a pseudoscalar. Consistency of the transformations $\phi \rightarrow -\phi$, $E \rightarrow -E$ and $\hat{c} \cdot \vec{\tau}_\perp \rightarrow -\hat{c} \cdot \vec{\tau}_\perp$ under $\sigma_v(\text{bc})$ then requires that $\tau_\perp(-\phi, -E) = -\hat{c} \cdot \vec{\tau}_\perp = -\tau_\perp(\phi, E)$. One can also show that the cross product of a vector and a pseudovector transforms as a vector, and so $\hat{m} \times \hat{c}$ is a vector. This implies that $(\hat{m} \times \hat{c}) \cdot \vec{\tau}_\parallel$ transforms as a pseudoscalar so that $(\hat{m} \times \hat{c}) \cdot \vec{\tau}_\parallel \rightarrow -(\hat{m} \times \hat{c}) \cdot \vec{\tau}_\parallel$ under $\sigma_v(\text{bc})$, and therefore $\tau_\parallel(-\phi, -E) = -\tau_\parallel(\phi, E)$. We have considered only torques linear in $E$ so that the symmetry requirement becomes $\tau_{\perp(\parallel)}(-\phi, E) = \tau_{\perp(\parallel)}(\phi, E)$. Keeping only the terms in Eq. (S6) that comply with this symmetry requirement leaves

$$\tau_\parallel(\phi, E) = E(S_0 + S_1 \cos\phi + S_3 \cos 2\phi + S_5 \cos 3\phi + \ldots)$$
$$\tau_\perp(\phi, E) = E(A_0 + A_1 \cos\phi + A_3 \cos 2\phi + A_5 \cos 3\phi + \ldots)$$ . (S7)

The measured angular dependence discussed in the main text for $E$ along the a-axis can be fit accurately with just the low-order terms $S_1$, $A_0$, and $A_1$. Notably, we do not experimentally observe the term $S_0$, although it is allowed by symmetry.

For an electric field applied along the b-axis, applying $\sigma_v(\text{bc})$ to the device flips the projection of the magnetization along the b-axis direction, and leaves the electric field unchanged i.e. $\phi \rightarrow \pi - \phi$ and $E \rightarrow E$. From this, one can derive the symmetry constraints $\tau_{\perp(\parallel)}(\pi - \phi, E) = -\tau_{\perp(\parallel)}(\phi, E)$. Therefore the allowed angular dependencies of the torques for an electric field $E$ along the b-axis are

$$\tau_\parallel(\phi, E) = E(S_1 \cos\phi + S_4 \sin 2\phi + S_5 \cos 3\phi + \ldots)$$
$$\tau_\perp(\phi, E) = E(A_1 \cos\phi + A_4 \sin 2\phi + A_5 \cos 3\phi + \ldots)$$ . (S8)

In this case, with $E$ along the b-axis, the lowest order terms ($S_1$ and $A_1$) dominate our measurements for both the symmetric and antisymmetric amplitudes, although better agreement is obtained when we include the coefficient $A_5$ as shown in Fig. S5.

**Supplementary Note 5: Higher harmonics in the ST-FMR angular dependence**

Based on the symmetry analysis in Supplementary Note 4, we may expect that the angular dependence of the in- and out-of-plane torques can be more general than $\tau_\perp = B + A\cos\phi$ and $\tau_\parallel = S\cos\phi$. We examined fits of our data to the most general symmetry-allowed Fourier expansion, up to the third harmonic. We find significant values for $A_5$, with the largest magnitudes occurring for current flowing close to the b-axis direction. Figure S5 shows

$V_A$ as a function of $\phi$ for two devices, along with fits to $\sin(2\phi-2\phi_0)[B+A\cos(\phi-\phi_0)]$ and $\sin(2\phi-2\phi_0)[B+A\cos(\phi-\phi_0)+C\cos(3\phi-3\phi_0)]$; the $\cos 3\phi$ term significantly improves the fit, corresponding to a non-zero value of $A_5$. We also find significant values for $S_5$, but $S_5/A_1$ is typically similar in magnitude to its value for our Pt/Py control device ($S_5/A_1 = -0.10\pm0.02$). All other coefficients up to the third harmonic, except for those discussed in the main text, are zero within our experimental uncertainty.

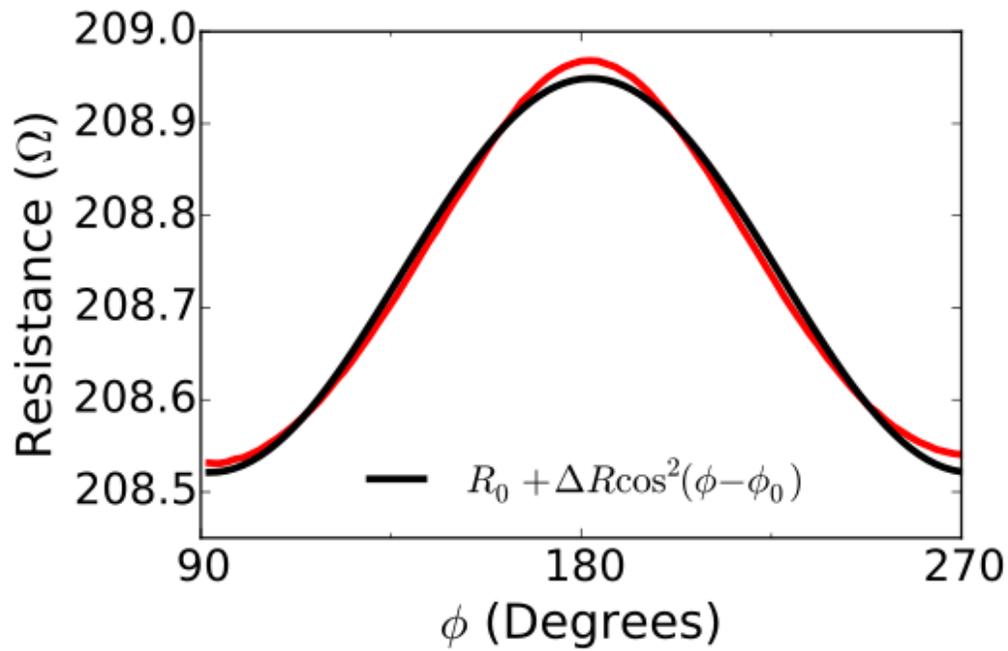

Figure S1: Resistance of Device 1 (red) as a function of applied in-plane magnetic field angle. Measurements are made in a Wheatstone bridge configuration with a static magnetic field of 0.08 T. The fit (black) is used to extract values of $dR/d\phi$.

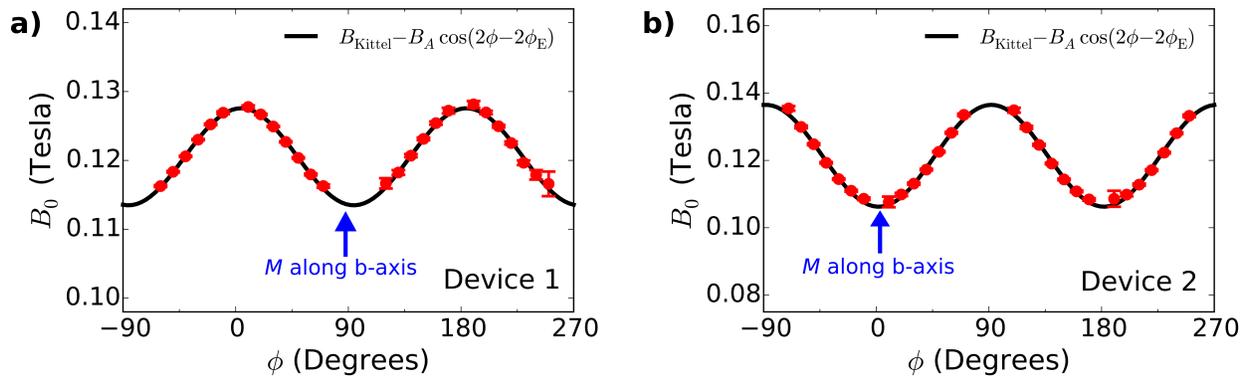

Figure S2: Ferromagnetic resonance field as a function of the in-plane magnetization angle for (a) Device 1 and (b) Device 2. The data are represented by red circles and the black lines are the indicated fits. In both cases the applied microwave frequency is 9 GHz and the power is 5 dBm. The blue arrows indicates the values of $\phi$ for which the magnetization lies along the b-axis.

| Device Number | $t$ (nm) ±0.3 nm | $l \times w$ (μm) ±0.2 μm | $\tau_B / \tau_A$ | $\tau_S / \tau_A$ | $\phi_{a\text{-}I}$ (Degrees) ±2° | $B_A$ (mT) | $\phi_{\text{Easy-I}} + 90°$ (Degrees) |
|---|---|---|---|---|---|---|---|
| 1 | 5.5 | 4.8 X 4 | 0.373(4) | 0.72(1) | -5 | 7.0(7) | 3.4(3) |
| 2 | 15.0 | 6 X 4 | 0.011(7) | 0.77(3) | 86 | 15.1(2) | 84.9(6) |
| 3 | 3.1 | 3.5 X 4 | -0.372(6) | 0.84(2) | -3 | 6.2(4) | 4.2(9) |
| 4 | 5.6 | 4 X 4 | -0.47(1) | 0.74(6) | -5 | 4.9(12) | 2(3) |
| 5 | 8.2 | 6 X 4 | 0.133(8) | 0.99(3) | 70 | 15.0(1) | 74.7(5) |
| 6 | 3.9 | 6 X 4 | 0.372(9) | 0.70(3) | -4 | 9.8(2) | 2.7(7) |
| 7 | 3.4 | 4 X 3 | 0.207(8) | 1.20(3) | 70 | 15.3(1) | 75.1(4) |
| 8 | 2.2 | 4 X 3 | 0.385(7) | 0.83(3) | -9 | 7.4(1) | -0.3(5) |
| 9 | 6.7 | 5 X 4 | 0.278(6) | 0.70(2) | 19 | 17.3(1) | 24.7(5) |
| 10 | 2.8 | 4 X 3 | 0.095(8) | 1.42(3) | 77 | 11.6(2) | 80.2(4) |
| 11 | 14.0 | 5 X 4 | -0.13(1) | 0.72(4) | -56 | 13.8(2) | -58(1) |
| 12 | 5.3 | 5 X 4 | -0.320(6) | 0.70(2) | -9 | 15.6(3) | -6.0(3) |
| 13 | 1.8 | 5 X 4 | -0.045(4) | 0.79(2) | 82 | 17.2(2) | 83.4(4) |
| 14 | 5.3 | 5 X 4 | 0.340(7) | 0.78(3) | -25 | 14.0(1) | -20.9(5) |
| 15 | 5.5 | 5 X 4 | 0.332(7) | 0.74(2) | -16 | 15.5(1) | -14.8(5) |
| Pt/Py | 6 | 10 X 5 | 0.000(4) | 1.79(2) | N/A | 4.2(2) | 85.5(8) |

Supplemental Table S1: Comparison of device parameters for the WTe$_2$/Py bilayers discussed in the main text and a Pt/Py control device. $t$ is the thickness of the WTe$_2$ or Pt, $l$ is the sample length, $w$ is the sample width, $\tau_B / \tau_A$ and $\tau_S / \tau_A$ are the torque ratios defined in the main text, $\phi_{a\text{-}I}$ is the angle between the a-axis and the applied current, $B_A$ is the anisotropy field within the sample plane (see Extended Data Figure S2), and $\phi_{\text{Easy-I}}$ is the angle of the magnetic easy axis with respect to the applied current.

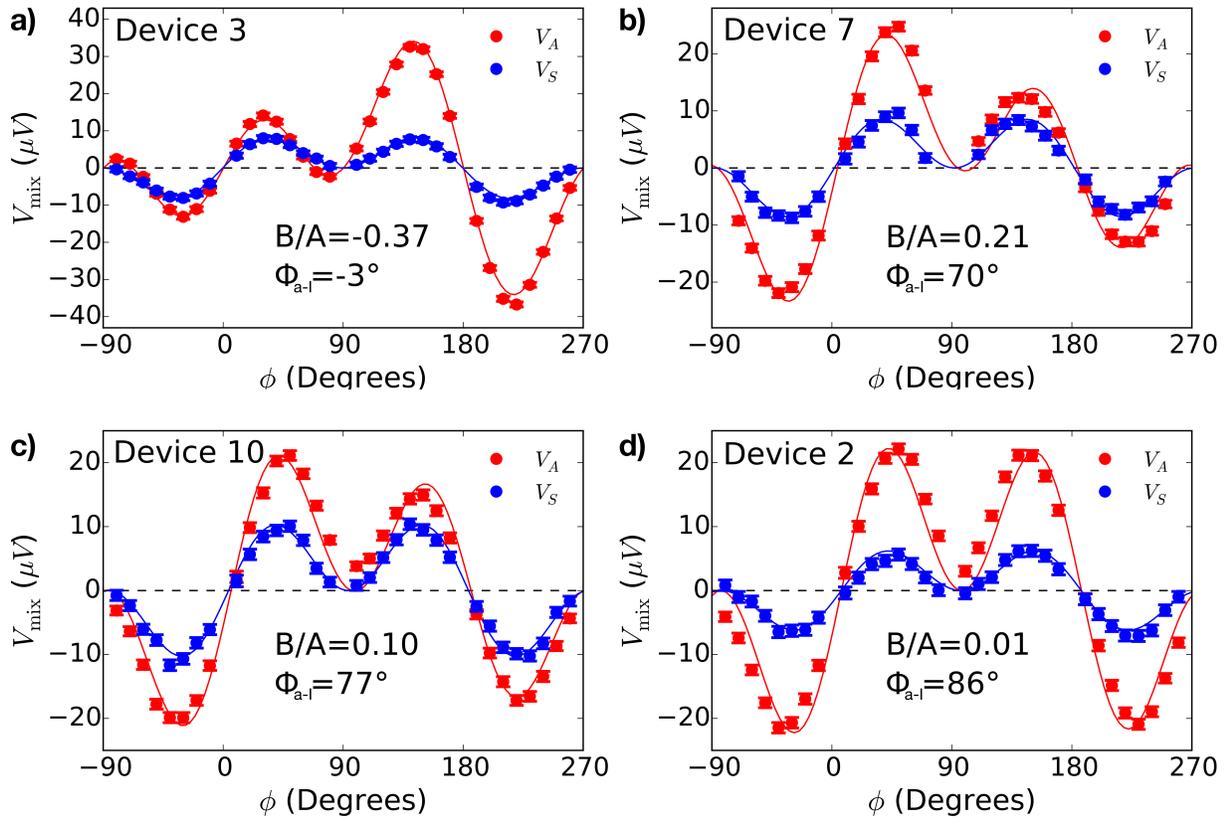

Figure S3: Plots of the symmetric (blue circles) and antisymmetric (red circles) components of the ST-FMR mixing voltage for (a) Device 3, (b) Device 7, (c) Device 10, and (d) Device 2. The current in Device 3 is applied approximately along the a-axis of the WTe$_2$, with the angle turning increasingly toward the b-axis for Devices 7, 10, and 2. The microwave frequency is 9 GHz and the microwave power is 5 dBm. The solid blue lines are fits of $S\sin(2\phi - 2\phi_0)\cos(\phi - \phi_0)$ to $V_S(\phi)$ and the solid red lines are fits of $\sin(2\phi - 2\phi_0)[B + A\cos(\phi - \phi_0)]$ to $V_A(\phi)$.

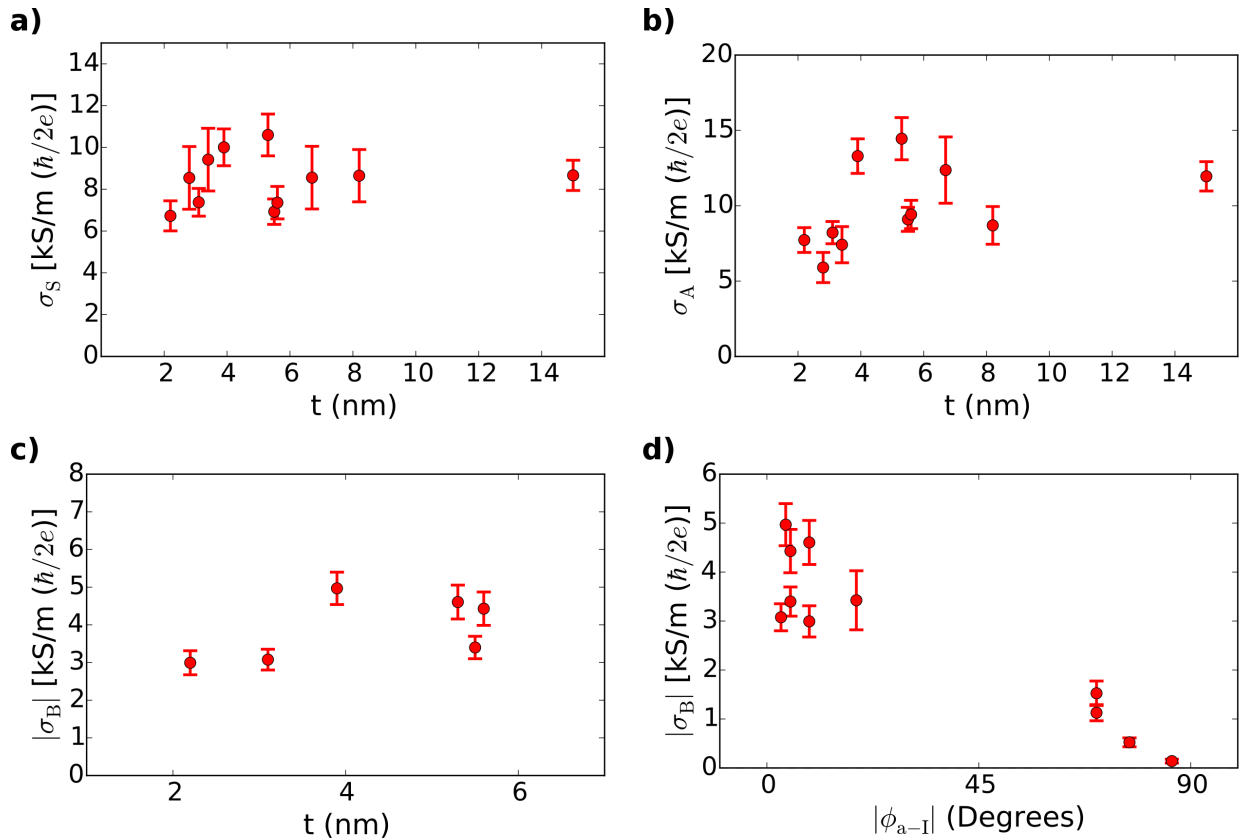

Figure S4: a) Torque conductivity $\sigma_S$ as a function of WTe$_2$ thickness for the 11 devices on which we used a vector network analyzer to perform fully-calibrated measurements. The current is applied at various angles to the WTe$_2$ a-axis. b) Torque conductivity $\sigma_A$ as a function of WTe$_2$ thickness for these 11 devices. c) Torque conductivity $|\sigma_B|$ as a function of WTe$_2$ thickness for 6 fully-calibrated devices with $|\phi_{a-I}| < 10°$. d) $|\sigma_B|$ as a function of $|\phi_{a-I}|$ for the 11 devices used in panels a) and b). For the data shown in panels a-d, the applied microwave power is 5 dBm, and the torque conductivities are averaged over the frequency range 8-11 GHz.

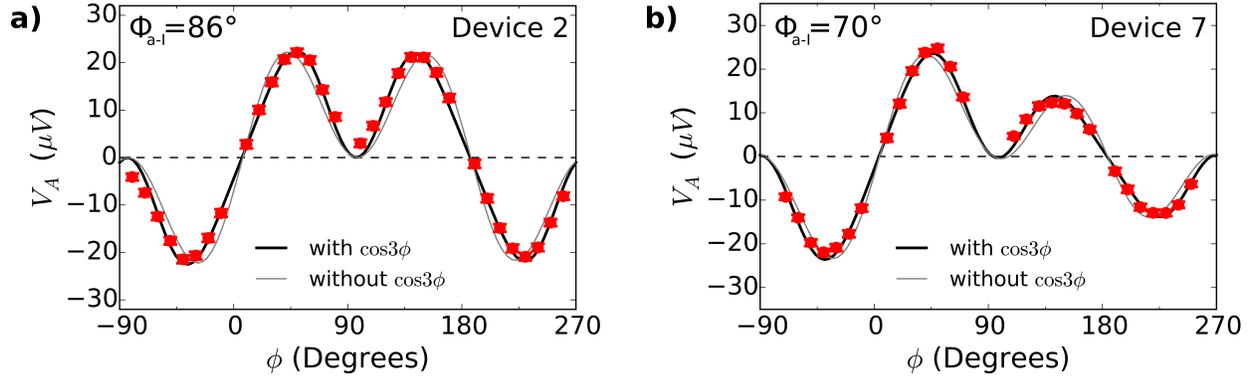

Figure S5: Plots of the antisymmetric part of the mixing voltage (red circles) versus the in-plane magnetization angle for (a) Device 2 and (b) Device 7. The black lines show fits to $\sin(2\phi - 2\phi_0)\left[B + A\cos(\phi - \phi_0) + C\cos(3\phi - 3\phi_0)\right]$, giving $C/A = -0.24 \pm 0.01$ for Device 2 and $C/A = -0.20 \pm 0.01$ for Device 7. The light grey lines show fits to $\sin(2\phi - 2\phi_0)\left[B + A\cos(\phi - \phi_0)\right]$.

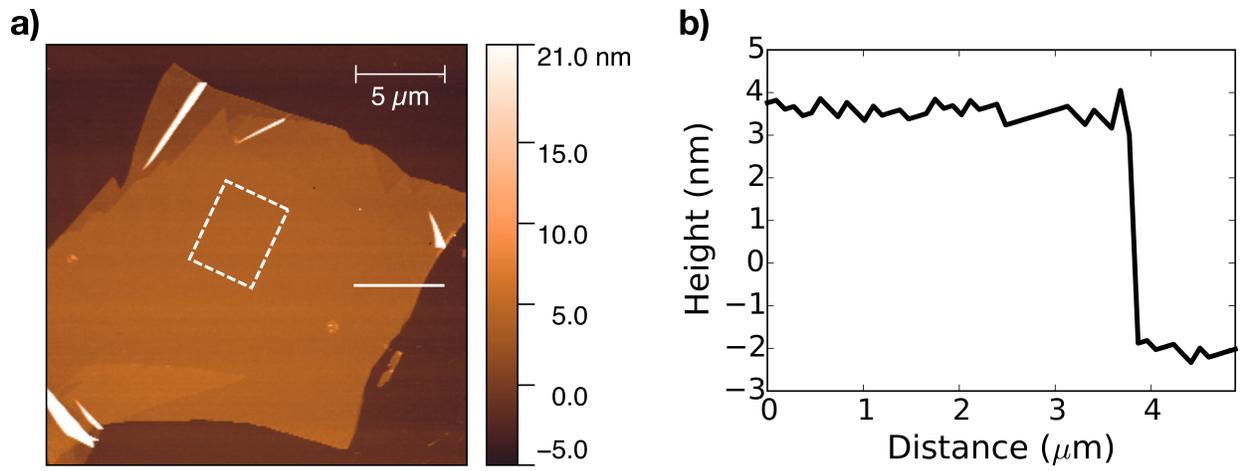

Figure S6: a) An atomic force microscopy image of the WTe$_2$ flake used for fabrication of Device 15 after deposition of the permalloy layer and aluminum oxide cap but before any lithographic processing. The active region used for the device (dashed white box) has a RMS surface roughness < 300 pm. b) A linecut [white line in (a)] from the edge of the WTe$_2$ flake, showing an average thickness of 5.5 nm.

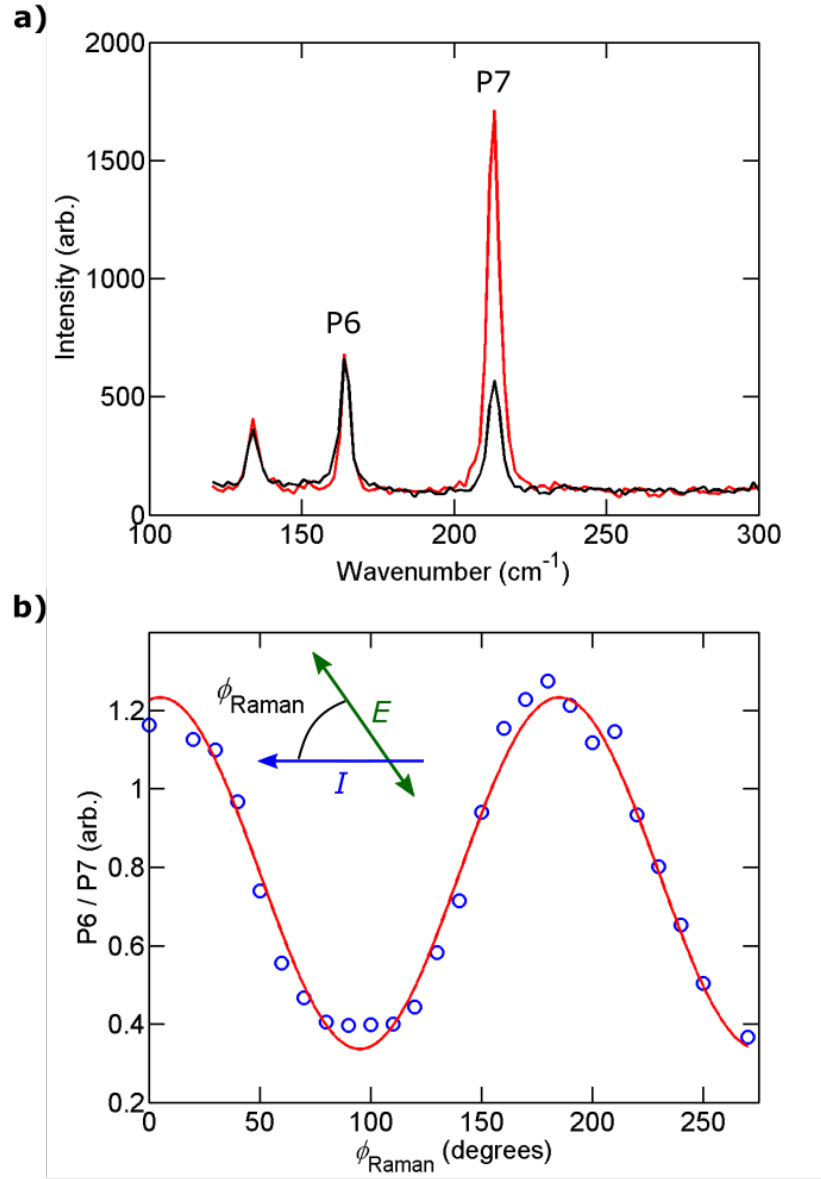

Figure S7: a) Polarized Raman spectra with the orientation of the electric field of the excitation, $E$, parallel to the WTe$_2$ a-axis (black) and parallel to the WTe$_2$ b-axis (red) for Device 4. Traces are normalized by the silicon substrate peak for ease of comparison (not shown). P6 = 165.7 cm$^{-1}$ and P7 = 211.3 cm$^{-1}$ (as defined in ref. S3) b) The ratio of intensities for P6/P7 (blue circles) plotted as a function of angle between the current (lithographically defined bar) direction and the linearly polarized Raman excitation as defined in the inset. The orientation of the WTe$_2$ a-axis is determined from the angle that maximizes the fit (red) to a $\cos^2(\phi_{Raman})$ type dependence (see ref. S3). The directions $\hat{b}$ and $-\hat{b}$ are not differentiated by Raman scattering.

**Supplemental References**